\documentclass[authoryear,12pt,3p]{jowarticle}

\usepackage[english]{babel}
\usepackage[utf8]{inputenc}
\usepackage{amsmath}
\usepackage{graphicx}
\usepackage[colorinlistoftodos]{todonotes}
\usepackage{natbib}

\makeatletter
\renewcommand\section{\@startsection{section}{1}{\z@}{-3.25ex plus -1ex minus -.2ex}{1.5ex plus .2ex}{\normalsize\bf}}
\renewcommand\subsection{\@startsection{subsection}{2}{\z@}{-3.25ex plus -1ex minus -.2ex}{1.5ex plus .2ex}{\normalsize\bf}}
\renewcommand\subsubsection{\@startsection{subsubsection}{3}{\z@}{-3.25ex plus -1ex minus -.2ex}{1.5ex plus .2ex}{\normalsize\bf}}
\makeatother

\begin{document}
\begin{frontmatter}
\title{The Natural Selection of Conservative Science}

\author{Cailin O'Connor}\ead{cailino@uci.edu}
\address{Department of Logic and Philosophy of Science \\ University of California, Irvine}

\date{\today}

\begin{abstract}
Social epistemologists have argued that high risk, high reward science has an important role to play in scientific communities.  Recently, though, it has also been argued that various scientific fields seem to be trending towards conservatism---the increasing production of what \citet{kuhn1970structure} would have called `normal science'.  This paper will explore a possible explanation for this sort of trend: that the process by which scientific research groups form, grow, and dissolve might be inherently hostile to high risk science.  In particular, I employ a paradigm developed by \citet{smaldino2016natural} that treats a scientific community as a population undergoing selection.  As will become clear, perhaps counter-intuitively this sort of process in some ways promotes high risk, high reward science.  But, as I will point out, high risk high reward science is, in general, the sort of thing that is hard to repeat.  While more conservative scientists will be able to train students capable of continuing their successful projects, and so create thriving lineages, successful risky science may not be the sort of thing one can easily pass on.  In such cases, the structure of scientific communities selects against high risk, high rewards projects.  More generally, this paper makes clear that there are at least two processes to consider in thinking about how incentives shape scientific communities---the process by which individual scientists make choices about their careers and research, and the selective process governing the formation of new research groups.  
\end{abstract}
\end{frontmatter}

\section{Introduction}

Social epistemologists have argued that high risk, high reward science has an important role to play in scientific communities.  Such science may open new and exciting areas of exploration. Those responsible for such ventures may go on to found new fields, often to great acclaim.  Nonetheless, some authors have recently argued that various scientific fields seem to be trending towards conservatism---the increasing production of what Kuhn would have called `normal science' \citep{luukkonen2012conservatism}. Funding bodies like NIH have recently introduced grants intended to reverse this perceived trend.\footnote{More details of the NIH High Risk, High Reward Research Program are available at their website: https://commonfund.nih.gov/highrisk.}  \citet{stanford2015unconceived} outlines some of the causal factors potentially implicated in increasing scientific conservativism, including the impact of funding bodies, the role of peer review in publication, the increasing age of new primary investigators (PIs), and severe competition for academic positions which might discourage risk-taking.\footnote{\citet{currie2017} also discusses these and other aspects of science that promote conservatism.  \citet{kummerfeld2015conservatism} argue that there is a free rider problem where every scientist should want to avoid risky experimentation and let others do it.  The implication is that rational choice considerations should drive scientists towards conservative projects}  

This paper will explore a different possible cause for trends toward conservatism in science.  As I will argue, the process by which scientific research groups form, produce, and dissolve might be inherently hostile to high risk, high reward science.  In particular, I employ a paradigm developed by \citet{smaldino2016natural} that models a scientific community as a population undergoing a type of selection.  The idea is that lab groups have different practices which contribute to their success, and which thus influence the likelihood that students from these labs end up forming their own labs.  On the reasonable assumption that students tend to adopt some of the practices of their academic advisors, we then have the three key ingredients of a Darwinian process---variation, selection on this variation, and heritability of variation \citep{godfrey2009darwinian}.

I will first consider the possibility that the significant chance of failure for high risk science means that fewer scientific mavericks will influence the next crop of successful scientists.  While this is the case, in fact the sort of selection process modeled here actually gives an advantage to high risk, high reward science.  Since labs that take risks tend to have a greater variance in terms of their success, many of them will not tend to place students.  But the most successful of them will be so successful that they will have an out-sized impact on the scientific community.  This is especially true when academic competition is fierce, counter to the suggestion from \citet{stanford2015unconceived} that competitive environments dampen risk-taking in science.

As I will point out, though, successful risky science is often hard to repeat.  While more conservative scientists will be able to train students capable of continuing their successful projects, and so create thriving lineages, successful risky science may not be the sort of thing one can easily pass on.  In biological terms, success for risk-takers may not be as heritable as for conservative science.  Inasmuch as this is right, I will show how the structure of scientific communities will select against high risk, high reward projects.  

I conclude by discussing the relevance of these simple models to our understanding of scientific communities and to scientific progress.  While idealized models like those presented here can play many roles in argumentation\footnote{See \citet{downes1992importance,grune2013appraising, o2016black}.}, the ones in this paper fall most squarely under `how-possibly' modeling and also as tools for thought experiments and aids to reasoning. They also give a more general proof-of-possibility by illustrating how credit incentives in science can work at two different levels.  Incentives influence the choices of individual scientists, but also the selection of methods and practices at the community level.  As I will highlight, sometimes influences at these two levels will have different effects, meaning that in assessing potential interventions on scientific communities both levels should be taken into account.

In section \ref{sec:riskconserve}, I discuss the distinction between high risk, high reward science, and conservative science.  Although this divide is a coarse one, it is nonetheless a useful distinction for the current exploration.  In section \ref{sec:model}, I introduce the framework developed by \citet{smaldino2016natural} and describe, in detail, the model presented in this paper.  Section \ref{sec:results} describes results from these models, showing, in particular, how the risk/reward trade-off will help determine whether high risk, high reward science is `selected' in academic communities, and also demonstrating how the inherent uncertainty in risky science may make it the sort of thing that cannot be effectively passed on.  Section \ref{sec:conclusion} includes a discussion of what sorts of epistemic roles these models can play, and what they tell us about risky science.

\section{Risky Science and Conservative Science}
\label{sec:riskconserve}

The distinction between normal and revolutionary science goes back to \citet{kuhn1970structure}, who contrasted work that proceeds incrementally within a well established paradigm with work that seeks to develop a new paradigm.  In fact there is not always a clear dividing line between conservative science and innovative, groundbreaking, or maverick-y science. This said, in the extreme cases we can distinguish science that takes small bites, moves along well worn pathways, and follows establishment rules from science that bucks establishment trends and attempts ambitious projects.  Examples of the first sort might include the herpetologist who painstakingly documents features of a new species of newt in the rain forests of Costa Rica, or the neurobiologist who knocks out a series of genes related to amygdala development and documents the effects on developing chick embryos, or the economist who analyzes one more variant of a model of household bargaining.  In the second category we might include those who attempt to outline new unifying physical theories, or who suggest spraying reflective particles out of hot air balloons to prevent global warming, or who develop new hypotheses on the origins of life.  In between we can identify examples of science that is more or less groundbreaking or innovative.

Sometimes conservative science pays off in big ways, both epistemically and in terms of credit for scientists.  A classic example comes from Alexander Fleming's work on Penicillin notatum.  Fleming was absorbed in routine work in bacteriology when an accidental contamination of old petri dishes with mold led to the discovery that penicillin could inhibit bacterial growth \citep{hare1982new}.  The discovery had massive impacts on public health.  He has since been widely quoted as saying, ``When I woke up just after dawn on September 28, 1928, I certainly didn’t plan to revolutionize all medicine by discovering the world’s first antibiotic, or bacteria killer. But I guess that was exactly what I did.''\footnote{See \citet{Markel2013}.}  This discovery, of course, also had massive impacts on Fleming's career. More often, conservative science is a source of dependable incremental progress.  Conservative science tends to be fairly dependable, as well, in terms of whether the scientists who do it receive credit for their work.  The economist developing a new model of household bargaining is fairly likely to be able to publish their findings in a decent journal, and to use this publication towards tenure and promotion.

Not all sorts of science are dependable in this way. Some projects carry more inherent risk, in the sense that they may fail to generate successful or publishable findings.  Projects that are very innovative or novel carry further risk in that the investigator may end up labeled as an outsider (or even a quack), with all the associated career detriments.  The unifying E8 theory, for example, developed by surfer-physicist Garrett Lisi \citep{lisi2007exceptionally}, is enormously novel and ambitious, but not long after its introduction was rejected by much of the physics community
\citep{Lisiart}.  

On the other hand, it has been widely argued that the sorts of innovative projects that carry high risks of failure are often those that, when successful, have the greatest scientific impact. \citet{stanford2015unconceived}, for example, has argued that transformative, risky science may play a special role in theory change in that it is more likely to yield previously unconceived theories that might lie outside of the scientific mainstream.  \citet{thoma2015epistemic} uses `epistemic landscape' models to argue that the presence of scientists who do innovative research may improve the outcomes of an entire community.\footnote{\citet{weisberg2009epistemic} first developed epistemic landscape models to make this argument, but see \citet{thoma2015epistemic, alexander2015epistemic}.} \citet{currie2017} argues that certain areas of inquiry, such as into existential risks (those that threaten the human race), require especially innovative methods.  Likewise, if E8 had been adopted as a unifying physical theory, it would have led to a revolution in physics, and massive personal success for Lisi.  These are the sorts of projects that funding bodies like NIH seek to promote.  Going forward, I will sometimes refer to such projects as `risky', but this should be taken as shorthand for high risk, high reward science.\footnote{No one is interested in science that is risky in the sense of being likely to fail, but with no chance of making significant epistemic progress.}

So there are reasons to think that pursuing at least some amount of high risk, high reward science is important to scientific progress, and can lead to great personal reward for scientists.  But, as briefly described in the introduction, the current structure of science may promote inherently conservative work \citep{stanford2015unconceived,luukkonen2012conservatism}.  I now explore a new potential cause of conservativism in science: selection.  The idea, as will become clear, is that the processes by which new PIs are hired and form labs may select against scientists who engage in high risk, high reward science.

\section{Model}
\label{sec:model}

This paper borrows a framework developed by \citet{smaldino2016natural}.\footnote{\citet{mcelreath2015replication} give an earlier version of this model, but the one I present here is closest to \citet{smaldino2016natural}.  \citet{holman2017experimentation} also describe a selective model of scientific communities where industry influences the prevalence of methodologies favorable to their interests by funding labs that use these methodologies.}  Their model simulates a community of scientists with some number of labs, $N$.  Each time step in their simulations involves two stages: science and evolution.  During the science stage, each lab completes a project with some probability.  The likelihood that new research is produced and published depend on the methodological practices of the lab in question.  \citet{smaldino2016natural} assume that different practices will be more or less likely to generate positive results, and more or less likely to identify which of these results are actually false positives.  Labs that generate and publish new results receive some level of credit for their work, which accumulates over the course of the simulation.  

During the second stage of each time step, a partially selective process determines how the make up of the scientific community shifts.  First, some number of labs, $d$, are sampled, and the eldest of these is selected to `die off'.  This represents something like the retirement of a PI in a real community.  Then a new set of labs (also of size $d$) are sampled, and the lab with the highest accumulated credit `replicates', the idea being that students from successful labs are more likely to be hired by peer institutions and found labs of their own.\footnote{This is similar to the Moran process in biology, described by \citet{moran1958random}.  This process involves individuals in a finite population, where at each time step one individual is randomly chosen for death, and one is chosen for reproduction based on their relative fitnesses.}  \citet{smaldino2016natural} assume that students will not necessarily form labs exactly like their mentors, but that there is some influence that extends between advisee and advisor.  The new lab is like the old one modulo some `mutation' of the practices mentioned above.

I will present a novel version of this model adapted to investigate the selection of risky science. Variation between labs in this model will involve differential tendencies towards risk-taking in science. In particular, I assume that some labs are inherently conservative, and as
a result have a dependable probability of success, $p_C$, and a set credit payoff should success
occur, $u_C$. Some labs are inclined towards risky science, and for these labs the expected
payoff given a success is $u_R$, where it is assumed that $u_R > u_C$, or that across the community
the success of a risky project will yield more credit than for a conservative project.\footnote{As noted with the penicillin case in section 2, this will not always be a good assumption, but it is a simplification intended to
capture the general features of the science at hand.  Adding a rare large payoff for conservative science to the model will not significantly influence results.}  The probability that a risky project succeeds, however, will vary from lab to lab. This is meant to capture a situation where some risk takers happen to develop revolutionary ideas and projects, which tend to pay off over the course of their scientific careers, whereas others
languish. The probability that each risk-taking lab completes a successful project, $p^i_R$, is
selected randomly. 

The distribution I use ensures that 1) low success rates are common and higher ones less so and 2) there is an upper limit on the success rate of risky projects so that for all labs $p^i_
R < p_C$ (conservative projects succeed more often than all risky projects). This is done by selecting a random number $x$ from $[0,1]$ and then taking $p^i_R = \text{max}(0, cx^2-f)$.  On the domain $[0,1]$ the function $cx^2-f$ is monotonically increasing and convex (meaning that higher values are increasingly less likely).  The constants $c$ and $f$ act as parameters which control how swiftly the function increases ($c$) and how many values map to zero probability of success ($f$).  The value at $x = 1$ is the maximum $p^i_R$ that can be obtained.  When $c = .4$ and $f = .02$, for example, the maximum value for $p^i_R = .38$.  For the parameter values I consider here, there are always some significant percentage of risk-taking labs that have no chance of success, and the chances are always significantly less than conservative labs.  This is obviously an arbitrary method for assignment of success rates, but any function with a similar general character should generate qualitatively similar results.

Simulations of this model proceed as follows.  First $N = 100$ labs are initialized to be either risky or conservative.  For simplicity I will always assume that it is equally likely a lab is conservative or risk-taking at the start of simulation.  For each risky lab, a characteristic success rate ($p^i_R$) is chosen.  Then, as in the version of the model presented by \citet{smaldino2016natural}, successive stages of science and evolution proceed.  Over time a lab accumulates credit.  At each stage the oldest lab from a random sample of size $d$ `dies', and the most successful of the second sample of $d$ labs replaces it.  Since all conservative labs have the same success rate and payoff, students of conservative labs inherit the properties of their parent lab perfectly.  They do conservative science with chance of success $p_C$ and credit $u_C$.  For risk-taking labs, I explore the possibility that there might be something difficult to repeat about risky science, so that although students of risk-takers also take risks, their success rate is chosen anew with some probability.  In particular, I define parameter $t$ as the probability that students inherit the success rate of their adviser.  When $t = 1$, the success of risk-taking, like the success of conservative science, is completely heritable.  When $t=0$, there is no correlation between the success rate of a risk-taking advisor and their students.\footnote{One variation of the model would consider the possibility that students of conservative scientists might themselves be risk-takers or vice versa.  This is certainly realistic.  While this change would alter the results, it would not significantly influence the main conclusions drawn from these models, which are about the environments under which risk-taking and conservatism come under positive selection pressure.}

Before presenting results from these models, a few comments are in order.  The first regards the interpretation of `success' or `payoff' in these models.  Many philosophers of science have gained insight into the behaviors of scientists by assuming that they are part of a `credit economy' where they respond to academic credit incentives in the same way normal people respond to monetary incentives.\footnote{See \citet{merton1973sociology} for more on the credit motives of scientists.  \citet{kitcher1990division} presented an early model of this sort.  For more recent examples see \citet{heesen2017communism,bright2017fraud}. \citet{zollman2018credit} discusses whether and when credit motives are a good thing for science.} Like `utility' itself, credit in this sense is not a well-defined notion (except inasmuch as we might define it as the-thing-sought-by-scientists). It is meant to capture something like high social status within a discipline, fame, attention, citations, awards, and attendant monetary/professional benefits. The model presented here likewise assumes that labs accumulate credit of this sort, and that this accumulation helps determine which labs manage to place students.  Notice, though, that while credit to scientists often is associated with scientific advancement due to their work, this is not always the case.  In other words, high risk, high reward science as modeled here may not completely overlap with the sorts of revolutionary science that authors like \citet{stanford2015unconceived,thoma2015epistemic,currie2017} have championed.  The models apply best to cases where credit and scientific advancement are associated.

One further note.  As I argued in section \ref{sec:riskconserve}, the dividing line between `conservative' and `high risk-high reward' science is not always a clear one.  In order to study how selective forces might act on conservatism in science I have operationalized these concepts.  In particular, conservative science in the models here is science that leads to steady, dependable credit rewards for those who pursue it.  High risk, high reward science leads to a lower chance of receiving a great deal of credit.  Given the complexity of the phenomenon modeled, there will be cases in science where this operationalization, and thus the applicability of the models, have a better or worse fit.\footnote{Code for the model described in this paper is available at the URL: https://github.com/cailinmeister/Risky-Science}

\section{Results}
\label{sec:results}

For all the simulations reported here, the success rate for conservative labs was $p_C = .8$ and the credit payoff for a successful conservative project was $u_C = 1$.\footnote{It is not important that these parameters were not varied since the relationship between these values for risky versus conservative labs is what determines whether science in these models tends towards conservatism or risk-taking.}  Payoff for risky science, $u_R$, was varied from $2$ to $20$.  Parameter $c$, which controlled how likely it was for risky science to pay off (higher $c$ corresponding to more risk-taking labs with high characteristic chance of success), was varied from $.1$ to $.4$.\footnote{I kept $f = .02$, meaning that the percentage of risk-taking labs that were never successful ranged from about 25-50\% depending on $c$.}  I varied $t$, the probability that risk-taking labs inherit the success rate of their parent lab, from 0 to 1. For each set of parameter values, 1000 trials were run, each with 1000 rounds of science and evolution.

At the end of each simulation, it was noted whether the remaining labs were all conservative, all risk-taking, or whether there was some mix of both.  The nature of this simulation is such that with enough time all labs will either be conservative or risk-taking.  This is because the selection process is probabilistic, and eventually this stochasticity will mean that one of the strategies will die out.  Once this happens it will be impossible for the strategy to re-enter the population since there is no mutation from conservative to risk-taking or vice versa.  This is to say that `both' is a temporary state in the model, but it is still meaningful because in simulations where both strategies are present after 1000 rounds, we know there is not strong selection for either strategy.

\begin{figure}
\centering
\includegraphics[width=.8\textwidth]{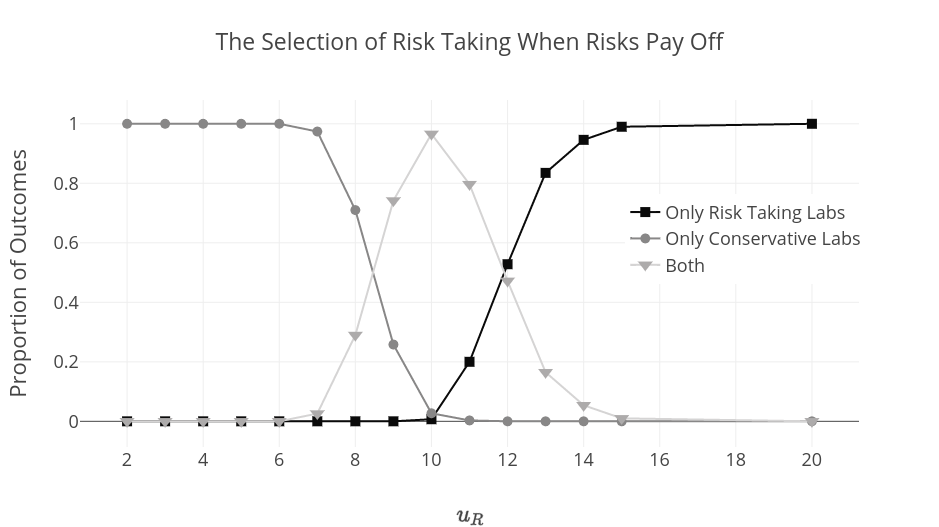}
\caption{Proportions of simulations that select risk-taking, conservative science, or both as the payoffs for risk-taking increase. For results displayed in this figure $c = .2$, $d = 10$, and $t = 0$.}
\label{fig:smeh}
\end{figure}

Unsurprisingly, across all simulations risk-taking was more likely to flourish when it generated more credit.  Figure \ref{fig:smeh} shows the proportion of simulations that generated risk-taking, conservativism or a mix of both as $u_R$ increases.  As is evident, the larger the payoff when risky science succeeds, the more likely it is that the selection process in the community will select for risk-taking. Although this figure represents one set of parameters, for this, and all other results described, trends were stable across all parameters.  An obvious take-away, though not a surprising one, is that if we wish to promote high risk science, the rewards for success should be increased relative to rewards for conservative science.     Notice that in this model, increasing credit rewards do not incentivize scientists to switch to risky projects.   Rather this model illustrates the ways that the processes by which new labs are formed will select for risky science when it generates a lot of credit, apart from the individual preferences and decisions of scientists. (In this case, of course, the intervention that should promote high risk, high reward science via selection is the same as the intervention that should incentivize credit motivated scientists to switch to risky projects.)\footnote{The question of what such an intervention would look like in this case is actually quite a tricky one.  How do we identify which science was high risk, high reward in the first place?  How do we give more credit to those scientists engaged in it?  Answering these questions is outside the scope of this paper, but the reader should be aware of difficulties here.  Thanks to an anonymous referee for raising this point.}

In addition, the greater the chances that risky science pays off (the higher $c$ is), the greater the probability it is selected for.  I will not present figures illustrating this result since it is so unsurprising.  Notice, though, that two of the causes of conservatism named by \citet{stanford2015unconceived}---low funding support and conservatism in peer review---will influence something like $c$.  If high risk projects are unlikely to get support and are hard to publish, labs engaging in risky science are less likely to gain credit successes.  Again, these factors should influence choices by scientists who predict that risky projects will be unlikely to get funding/be published, but they should also influence the selective processes by which new labs are formed.  And, again, in this case, the same incentives should promote high risk, high reward science via both pathways. 

Taken together these two results---regarding the payoff and probability of success for risky science---indicate that epistemic communities might tend towards conservatism when those engaged in risky science do not garner the success necessary to pass on their methods via students.  Despite the fact that the two trends just described are unsurprising, there is an aspect of these results that is unexpected.  One might think that what does the work in determining whether risky or conservative science tends to be selected here are the relative expected payoffs of the two scientific strategies.  For conservative science, the chance of success is always .8 and the credit payoff is 1, so the expected payoff on any round is .8 across the board.  For risky science, the expected payoffs across labs must take into account the variance in their success rates.  This expected payoff across labs will be the payoff for risky science, $u_R$ integrated over the various possible chances of success (at least at the start of simulation).\footnote{In other words, it will be $\int_{x=0}^1 u_R *\text{max}(0, cx^2-f)dx$ (see footnote 9 . ).} But when the expected payoffs for risky and conservative science are equal, these simulations select risky science.  In fact, they select risky science even when its expected payoff (in this sense) is much lower than that for conservative science.\footnote{Just to give one example, consider the results in figure \ref{fig:smeh}.  When $u_R = 10$, conservative and risky science are selected equally often, but at this point the expected payoff to risky science integrated over possible chances of success is $.5$ (compared to $.8$ for conservative science).  When $u_R$ is nearly 16, the expected payoffs are equal, but at this point risky science is selected in all simulations.}

Why would this be?  Labs that engage in risky science, remember, also generate higher payoffs when they do succeed.  This means that the very best risk-taking labs will have enormous capacity to generate credit, and so will tend to have accumulated the most payoff of any labs in the community. In other words, the credit variance of risk-taking labs is much higher, meaning that at the very top of lab performance there will be more risk-takers.  At the very bottom there will be more risk-takers too, but the high performers matter most in this sort of selective environment.\footnote{\citet{smaldino2016natural} find something similar with respect to the selection of low powered experiments.  Labs that put less effort into preventing false positives also have more time to generate new work.  Even if there is a good chance their work will fail to replicate, and even if there are consequences when this happens, there will be some labs that by chance never face these consequences.  These will generate the most credit of all labs and their methods will tend to spread.}  They are the ones who found lineages of successful students and spread risk-taking practices.  So, while the risks associated with maverick-y science can indeed impair its spread, the process by which scientific labs are formed, produce, and die actually seems to promote the spread of high risk, high reward science.

In the introduction, I briefly mentioned a suggestion from \citet{stanford2015unconceived} that increases in the severity of competition for scientific positions are likely a factor promoting the rise of conservative science.  The models here do not support this particular claim.  One can think of the size of the sample from which labs are chosen to die and replicate, $d$, as determining the strength of selection in these models.  At the extremes, if $d = 1$ death and replication are always completely random.  If $d = 100$ on the other hand, the oldest lab always dies and the most successful one always replicates, meaning there is no randomness, but a strong selective force instead.  As the strength of selection in these models increases, perhaps surprisingly, the emergence of risk-taking becomes \emph{more} common, not less.

Why?  As the strength of selection increases in this way, it is increasingly the case that only the very top performing labs will ever replicate.  Labs with high payoffs, but not the very peak payoffs, no longer make the cut.  And, as discussed, because risk-taking labs have higher variance in success, the very top performers will tend to be risk-takers.  This is the same general reason that risk-taking derives a relative benefit in these selective processes.  The further observation is that the more competitive the environment, the more the process will tend to select high risk, high reward science. 

\begin{figure}
\centering
\includegraphics[width=.8\textwidth]{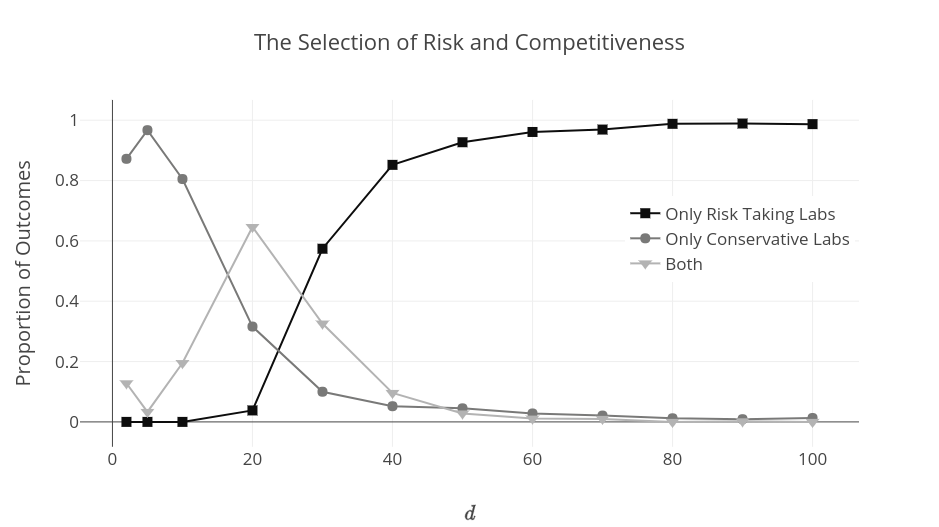}
\caption{Proportions of simulations that select risk-taking, conservative science, or both as selection pressure increases. For all results displayed in this figure $c = .2$, $t = .5$, and $u_R = 5$.}
\label{fig:Riskcompet}
\end{figure}

Figure \ref{fig:Riskcompet} shows this trend.  On the x-axis, the size of the samples from which labs are chosen to die and replicate increases.  As is evident, as this strength of selection increases, risk-taking is selected for with increasing frequency.\footnote{One may notice that on the far left of the plot, when $d = 2$, the proportion of simulations with both sorts of strategies at the end increases and the proportion of conservative labs goes down.  This is because when $d = 2$ there is very little selective pressure in this model, so the chances that either strategy dies out at any stage decrease.  When $d = 1$, so that there is no selection pressure at all, 99\% of simulations end with both strategies present.}  Given the simplicity of this model, it would be premature to take this as definitive evidence that highly competitive academic environments promote high risk, high reward science.\footnote{It would be especially premature because this model does not capture the fact that young investigators often must publish on their own in a short period of time in order to get a job.  This might increase selection pressure for projects that are more likely to yield a dependable success given a limited amount of time.}  But this said, these simulations do suggest that the connection which \citet{stanford2015unconceived} draws between competitive academic environments and conservatism may be overly hasty.  Even if competitive environments promote conservative decisions making by individual scientists, they seem to have the opposite effect on selection of conservatism.  Notice that this is a case where the process of change via selection on labs and the process of change via choice by scientists may pull apart.

At this point, the models would seem to suggest that in modern, highly competitive academic communities even when risk-taking is not a good bet for an individual scientist (because the expected payoffs are lower than for conservative science) it should nonetheless tend to spread as the students of the most successful labs found their own.  Let us now explore, though, the possibility that one reason scientific communities might trend towards conservatism could have to do with the heritability of risky versus conservative strategies in science.  Successful conservative science is often easy to pass down.  A reasonably bright student of the economist who builds models of household bargaining can likewise build models of household bargaining.  The student of the scientist documenting details of salamander ecology can likewise document salamander ecology.  It is less clear that the student of Einstein can replicate his successes.  To capture this feature, I assume that when a risky lab replicates it sometimes receives a new characteristic probability of success.  The idea is that the student involved has chosen a new, risky project where the probability of success is not tied to the success of their advisor's project.

This lack of heritability does indeed inhibit the emergence of risky science in these models.  The reason is that in models where the success rate is heritable, the risk-taking labs that tend to do well are those that happen to have a high characteristic rate of success.  Their methods spread across the community as students with the same high rates of success form their own labs, publish, gain credit, and train their own students.  Notice that this also means that over the course of simulation risky science becomes less risky on average.  In models without heritability, a lab with a high success rate will generate successful students that get their own labs, but their success rates, on average, are no higher than other risk takers.  This prevents thriving academic lineages from spreading high-risk, high-reward science throughout the community.

\begin{figure}
\centering
\includegraphics[width=.8\textwidth]{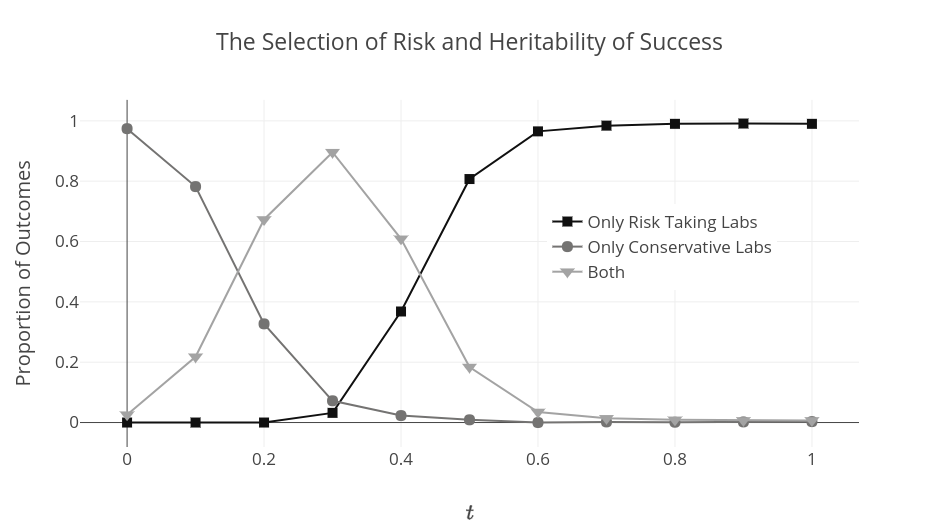}
\caption{Proportions of simulations that select risk-taking, conservative science, or both as the heritability of payoffs to risk-taking increases.  For these results, $u_R = 7$, $d = 10$, and $c = .2$}
\label{fig:Riskheritability}
\end{figure}

Figure \ref{fig:Riskheritability} demonstrates this trend.  On the x-axis we have the heritability of the success rate of risk-taking labs, $t$.  As the heritability increases, it is increasingly likely that risk-taking emerges.  When the heritability of success is low, on the other hand, conservativism tends to dominate.  The take-away here is that inasmuch as it is difficult to transfer the success of a high risk, high reward project on to graduate students, scientific communities may trend towards conservative projects where successful methods can be passed on.

This conclusion about heritability complicates a claim made previously in this section: that simply increasing the payoffs for risky science should promote it.  The models without heritability make clear that things are not so straightforward.   PIs who get a lot of credit for their successful high-risk, high reward science may spread these practices through successful students. But if their students are not able to replicate their successes, the process stops there. 

\section{Conclusion}
\label{sec:conclusion}

The models presented here are highly simplified, even in comparison to those that inspired them in \citet{smaldino2016natural}. As 
\citet{weisberg2012simulation} points out, no model can capture every desideratum a modeler might have (such as being maximally simple and maximally realistic). The goal here was to favor simplicity over realism in order to make the models tractable, easy to understand, and to get a clear picture of the causal processes occurring in the simulations.   For this reason, it is appropriate to be explicit about the ways they can and cannot inform our understanding of epistemic communities, and the selection of conservative science.

I take these models to be doing three main things.  First, they provide an aid to reasoning and thought experiment.  Simulations are not strictly necessary to generate and defend a hypothesis such as `if the success of risky science tends to be less heritable, it may be less likely to spread'.  However, the simulations here extend human reasoning capacity in order to make clear that the logic of this hypothesis holds up.\footnote{\citet{farrell2010computational} defend the use of simulations as an aid to reasoning in psychology.  As they point out, generating simulations of a phenomenon forces scientists to be precise about what they are talking about, and to use shared language and assumptions.  They also help scientists avoid cognitive biases such as confirmation bias, and to avoid problems with limitations in human memory and computational capacity.}  In addition, the models challenge an argument that seems plausible and defensible---that academic competition should increase conservatism.  In this challenge, the models are instrumental as an aid to reasoning, since their outcome is surprising, even if it is not hard to grasp the logic of the result once it has been outlined.

The second role I take these models to be playing is in `how-possibly' explanation.  \citet{dray1957} was the first to draw a distinction between `how-possibly' explanations---intended to refute claims of impossibility---and `how-actually' explanations.\footnote{\citet{grune2013appraising} gives a good description of the use of models in `how-possibly' explanations.}  The models here make clear how considerations other than the rational choices of individual scientists can be enough to drive an epistemic community towards conservative science.  In particular, they make clear that heritability, a factor which often comes apart from credit rewards, can possibly influence the emergence of scientific strategies.  Given that funding bodies such as the NIH have explicitly set out to promote high risk, high reward science, it is appropriate to tune into all the sorts of features of a scientific community that might lead to scientists avoiding such research. The models here introduce the possibility that some of these features may derive from larger structures of scientific communities, and may require different sorts of interventions in the interest of promoting high risk, high reward science.  The third role the models play is connected to the second but more general.  They help illustrate how, in principle, selective forces in scientific communities can act separately from processes of change via choice by individual scientists.  This bolsters claims by \citet{smaldino2016natural,holman2017experimentation} that selection can importantly shape scientific communities.

Of course, these models do not capture aspects of scientific behavior derived from rational choice, either motivated by credit or by epistemic goals.  For example, \cite{currie2017} points out that risk-averse students may be unwilling to join labs engaged in risky projects, and that PIs may avoid risk-taking to protect the students and postdoc under their supervision.  To be clear, the models presented here are intended to complement such work with a clearer picture of the other features of scientific communities that influence risk-taking, not to supplant them.

\section*{Acknowledgements}
Many thanks to The Center for the Study of Existential Risk at Cambridge University for funding and hosting the workshop that led to this paper.  Special thanks for Adrian Currie and Hugh Price for organizing.  Thanks to Adrian Currie, Paul Smaldino, Remco Heesen, and an anonymous reviewer for comments on the work.  And thanks to participants in the Risk and the Culture of Science workshop and to Paul Smaldino for conversations that inspired this paper.  This work was funded by NSF grant 1535139.  Thanks for their support.

\bibliographystyle{mla}
\bibliography{Riskbib} 

\end{document}